\documentclass[10pt,aps,prl,twocolumn,floatfix,nofootinbib,superscriptaddress,longbibliography]{revtex4-2}
\usepackage{placeins}
\usepackage{bm,graphicx,mathrsfs,amsmath,amssymb,mathtools,makecell,bbm, amsthm,dsfont,color,times,txfonts,nicefrac,framed,enumitem,tikz,physics,wrapfig,amsfonts,lipsum,nicematrix}
\usepackage[most]{tcolorbox}
\usepackage[dvipsnames]{xcolor}
\newlist{todolist}{itemize}{2}
\setlist[todolist]{label=$\square$}
\usepackage{mathtools}
\usepackage{cancel}
\usepackage{pifont}
\usepackage{algorithm}
\usepackage{algpseudocode}
\usepackage[colorlinks,linkcolor=NavyBlue,urlcolor=NavyBlue, citecolor=NavyBlue]{hyperref}
\usetikzlibrary{calc}% add hypertext capabilities

\newcounter{globalboxcounter} % Does not reset with sections
\renewcommand{\theglobalboxcounter}{\arabic{globalboxcounter}}
\newenvironment{sproof}{%
  \proof}{\endproof}

\newtheorem*{res*}{Main result}

\renewcommand{\v}[1]{\ensuremath{\boldsymbol #1}}

\definecolor{equationcolor}{RGB}{222,94,100}

\newtcolorbox[auto counter]{mybox}[2][]{%
    breakable,
    enhanced,
    sharp corners,
    colback=white,
    colframe=black!50,
    colbacktitle=white,
    coltitle=black,
    fonttitle=\normalfont,
    boxrule=0.25mm,
    arc=0mm,
    width=\linewidth,
    boxsep=1mm,
    left=1mm,
    right=1mm,
    top=1mm,
    bottom=1mm,
    before skip=6pt,
    after skip=6pt,
    before upper=\strut,
    title={\centering\strut \textbf{Example} {\normalfont\theglobalboxcounter: #2}},
    #1
}

\newcommand{\deriv}[2]{\frac{\text{d}{#1}}{\text{d}{#2}}}
\newcommand{\dderiv}[3]{\frac{\text{d}^{#1}{#2}}{\text{d}{#3}^{#1}}}
\newcommand{\cov}{\operatorname{cov}}
\renewcommand{\var}{\operatorname{var}}

\newcommand{\ms}[1]{\textsf{#1}}

\usepackage{pifont}% http://ctan.org/pkg/pifont

\def\A{ {\ms A} }
\def\B{ {\ms B} }

\def\X{ {\ms X} }
\def\Y{ {\ms Y} }

\usepackage{array}

\begin{document}

\preprint{APS/123-QED}

\title{Anomalous Local Heat Capacity and Bipartite Entanglement}

\author{Jake Xuereb}
\affiliation{Vienna Center for Quantum Science and Technology, Atominstitut, TU Wien, 1020 Vienna, Austria}
\email{jake.xuereb@tuwien.ac.at}

\author{A. de Oliveira Junior}
\affiliation{Center for Macroscopic Quantum States (bigQ), Department of Physics, Technical University of Denmark, 2800 Kongens Lyngby, Denmark}
\email{alexssandredeoliveira@gmail.com}

\begin{abstract}
The heat capacity of a system quantifies how it energetically responds to changes in temperature at equilibrium. Whilst this quantity is positive and even additive for non-interacting systems, self-gravitating systems such as stars or subsystems of strongly interacting quantum systems are known to have negative or anomalous specific heat capacities. In this work, we investigate how the presence of entanglement at equilibrium can influence how an interacting system responds energetically to changes in temperature. We examine the local heat capacity of interacting quantum systems providing an analytical understanding for when anomalies occur. Most interestingly, we find a connection between local heat capacity anomalies and entanglement by deriving a separability bound based on the fluctuations of local and interaction energies. We illustrate our results with two examples (i) a nearest neighbour spin-1/2 chain and (ii) two coupled quantum harmonic oscillators. Lastly, we provide an information-theoretic formula connecting mutual information and athermality to the non-additivity of the heat capacity. Our results provide model-independent thermodynamic entanglement detection bounds and insight into the relationship between quantum correlations and the heat capacity of quantum systems. 
\end{abstract}

\maketitle

\begin{figure}[t]
    \centering
    \includegraphics{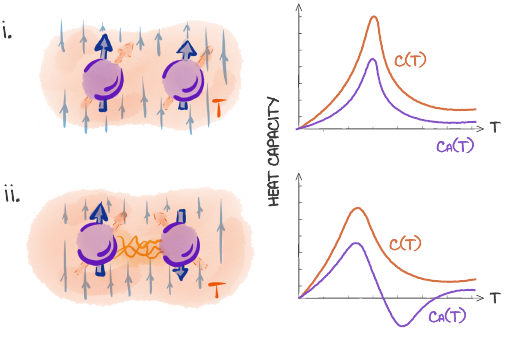}
    \caption{\emph{Anomalous heat capacity meets entanglement}. At thermal equilibrium, (i) the global heat capacity $C(T)$ (orange curve) of a noninteracting bipartite system is the sum of its local heat capacities, whereas (ii) in the presence of interactions, $C(T)$ need not to be equal to the sum of the local contributions. Although the global heat capacity remains nonnegative, a local heat capacity can become anomalous, i.e., negative, as illustrated by $C_\A(T)$ (purple curve). Here, we investigate when such anomalous local thermal responses are connected to entanglement.} 
    \label{F-general-idea}
\end{figure}

\emph{\textbf{Introduction.}} At equilibrium, the local kinetic energy of the molecules forming a star is anti-correlated with their potential energy under variations in temperature. That is, adding energy to the star increases its average energy as the particles move farther apart, whilst decreasing its average kinetic energy and hence the particles' speeds and temperature. This phenomenon underlies the gravothermal catastrophe in cosmology~\cite{bell_mnras_68,1970_thirring,lynden_bell_99}. Similar behaviour has also been documented in open quantum systems~\cite{ingold_09,campisi2010thermodynamic}, where the heat capacities of subsystems of strongly interacting quantum systems can become negative, giving rise to what is termed anomalous heat capacity~\cite{ingold_09}. Both phenomena teach us that interactions can change how systems at equilibrium respond to changes in temperature in surprising ways. In this work, we investigate \textit{whether the presence of quantum entanglement at equilibrium can influence how an interacting system responds to changes in temperature.} (Fig. \ref{F-general-idea})

The heat capacity~\cite{brukner2004macroscopicthermodynamicalwitnessesquantum,wiesniak_2005,brukner_08} and average energy at equilibrium~\cite{eisert_audenaert_02,dowling_04,anders_08,AndersWinter2008} have previously been used for entanglement detection, but existing approaches have either been model-dependent or required bounds on the average energy of separable states. In this work, focusing on interacting bipartite systems, we derive model-independent bounds on thermodynamic quantities whose violation detects entanglement at equilibrium. We study the energetic response
of the bare Hamiltonian to global changes in the equilibrium temperature, which we term the \textit{local heat capacity}. We find that these local heat capacities can take anomalous negative values providing a general analytical criterion for such anomalies. By deriving a local uncertainty relation~\cite{reid_88,reid_89,guhne_04,giovanetti_04} between local and interaction energy expectations we find an entanglement detection bound. Using this bound we show entanglement at equilibrium can at times necessitate local heat capacity anomalies, providing insight into the origin of the anomalies pointed out by Ingold et al.~\cite{ingold_09}. We showcase the bound by examining it numerically for a spin-$1/2$ chain. We then examine two coupled quantum harmonic oscillators, where despite the absence of anomalies, local heat capacities still carry signatures of entanglement. Finally, we use information-theoretic tools to systematically study the non-additivity of heat capacity in interacting systems. In short, our main technical result can be stated as follows. For $H=H_0+H_{\text{int}}$, with $H_0=H_\A+H_\B$ and $H_{\text{int}}=V_\A\otimes V_\B + W$, every separable equilibrium state satisfies
\begin{gather}
    \frac{C_\ms{A}(T) + C_\ms{B}(T)}{\beta^2} \geq \frac{\mathcal{O}_\beta^2}{4\var_\beta(V_\A\otimes V_\B)}+\cov_\beta(H_0,H_{\text{int}}).
\end{gather}
where $\mathcal{O}_\beta = \left(|\langle i [H_\ms{A},V_{\A}]\otimes V_{\B}\rangle_{\beta}| + |\langle V_{\A} \otimes i [H_\ms{B},V_{\B}]\rangle_{\beta}|\right)$. Violation of this bound detects entanglement and can, in scenarios we will explore, imply anomalous local heat capacities.

\emph{\textbf{Anomalous Local Heat Capacity}.} Heating a system in a Gibbs state of a fixed Hamiltonian cannot lower its total energy, since its heat capacity is non-negative. However, this is not always true locally for interacting composite systems. Surprisingly, the same temperature increase can lower the chosen energy of one, or even both, subsystems. To characterise this redistribution of the thermal response, we consider a bipartite system $\A\B$ with Hamiltonian $H=H_0+H_{\rm int}$, where $H_0:=H_{\ms A}+H_{\ms B}$ is the sum of fixed local Hamiltonians and $H_{\rm int}$ is the interaction Hamiltonian. The system is prepared in the Gibbs state $\rho_\beta=e^{-\beta H}/Z$, with partition function $Z:=\tr(e^{-\beta H})$, at inverse temperature $\beta:=T^{-1}$, where we set $k_{\rm B}=1$. Throughout, $\ms{X}\in\{\A,\B\}$, while $\ms Y$ denotes the complementary subsystem.

We are interested in how the energetic response to a change in temperature is distributed among the two subsystems and the interaction. For any temperature independent observable $O$, the Gibbs state satisfies the thermal response identity
\begin{equation}\label{Eq:thermal-response-identity}
    \deriv{\langle O\rangle_\beta}{T} = \beta^2 \cov_\beta(O,H),
\end{equation}
which follows by differentiating the Gibbs state. Throughout the paper, we denote thermal expectation values as \mbox{$\langle O\rangle_\beta=\tr(\rho_\beta O)$} and we use the covariance \mbox{$\cov_\beta(O_1,O_2) =\langle O_1 O_2\rangle_\beta - \langle O_1 \rangle_\beta\langle O_2\rangle_\beta$} and  the variance $\var_\beta(O)=\cov_\beta(O,O)$. It follows that for the total Hamiltonian, Eq.~\eqref{Eq:thermal-response-identity} gives the \emph{global} heat capacity:
\begin{align}\label{Eq:heat_cap_decomp}
  \hspace{-0.3cm}C(T) &= \deriv{\langle H \rangle_\beta}{T}=  \beta^2 \var_\beta(H) = C_\A(T)+C_\B(T)+C_{\rm int}(T),
\end{align}
where we define $C_{\ms{X}}(T):=\beta^2\cov_\beta(H_{\ms X},H)$ to be the \emph{local heat capacity} of the subsystem $\ms{X}$ and $C_{\rm int}(T):= \beta^2 \cov_\beta (H_{\rm int},H)$ is the interaction contribution. Thus, the global heat capacity is not, in general, the sum of the two local responses, although additivity is recovered in the non-interacting limit. Importantly, $C_\ms{X}(T)$ is not the heat capacity of an independently thermalised subsystem or the Hamiltonian-of-mean-force heat capacity discussed in the End Matter. It is the thermal susceptibility of the chosen local energy operator while the global Gibbs temperature is varied. As with any local observable, its expectation value may be evaluated using the reduced state $\rho_{\ms{X}}:=\tr_{\ms{Y}}(\rho_\beta)$, but $\rho_{\ms{X}}$ is generally not a Gibbs state of $H_\X$ at inverse temperature $\beta$. Operationally, because it is a susceptibility, the local heat capacity has the potential to be explored through calorimetric experiments~\cite{PhysRevLett.133.186704,PhysRevB.99.024413}. That is, if the subsystem is embedded in a larger, well-characterised macroscopic system, the change in energy of the subsystem can be measured through the macroscopic system without measuring in the energy eigenbasis of the interacting system.

\emph{When is the local heat capacity negative?} Unlike the global heat capacity, a local heat capacity is determined by a covariance and may have either sign. The decomposition of the total Hamiltonian, together with the bilinearity of the covariance, gives
\begin{equation}
    \frac{C_\X(T)}{\beta^2} = \var_\beta(H_\X)+\cov_\beta(H_\X,H_\Y) +\cov_\beta(H_\X,H_{\rm int}),
\end{equation}
where the first term is non-negative and quantifies local energy fluctuations, whereas the other two terms measure correlations between different energy contributions. Consequently, $C_\X(T)<0$ if and only if $-\cov_\beta(H_\X, H_\Y + H_{\rm int}) > \var_\beta H_\X$.

A negative local heat capacity therefore requires the local energy to be anti-correlated with the rest of the system strongly enough to overcome its own fluctuations. A simple illustration of this phenomenon is provided by the two-qubit XY model. Its Hamiltonian is \mbox{$H_{XY}=-h(\sigma^{A}_z + \sigma^{B}_z ) - J(\sigma^A_x\sigma^B_x + \sigma^A_y\sigma^B_y)$} with $h>0$ and $J=2h$. A closed-form calculation, presented in the Supplementary Material (SM), shows that the local energy of either qubit is anti-correlated with the remaining energy at all finite temperatures. This anti-correlation becomes sufficiently strong to overcome the local energy variance, thus making both local heat capacities negative, when $\beta h \!>\! \tfrac{1}{2}\operatorname{arccosh}(\tfrac{1+\sqrt{3}}{2})$. The ground state, governed by the interaction term, is the Bell state $\ket{\Psi^+}=2^{-\nicefrac{1}{2}}(\ket{01}+\ket{10})$, whose bare local energy vanishes. Heating initially populates the state $\ket{00}$, whose bare energy is $-h$ in each qubit, thus lowering both local energies even though the total energy increases as $E_{00} > E_{\Psi^+}$.

\emph{When can both local heat capacities be anomalous?} The exchange symmetry of the preceding model implies \mbox{$C_{\ms{A}}(T)=C_\B(T)$}, but the two responses need not have the same sign in a generic bipartite system. To determine when both responses are anomalous, we define $H_\Delta:=H_\A-H_\B$. Using the identity $\max\{a,b\} = \tfrac12(a+b+|a-b|)$ together with Eq.~\eqref{Eq:thermal-response-identity}, we find that 
\begin{equation}
\max_{\X\in\{\A,\B\}}C_\X(T) = \frac{\beta^2}{2}\qty[\cov_\beta(H_0,H_0+H_{\rm int})\!+\!|\cov_\beta(H_\Delta, H)|],
\end{equation}
and both local heat capacities are negative if and only if 
\begin{equation}\label{Eq:simultaneous-anomalies}
    -\cov_\beta(H_0,H_{\rm int}) > \var_\beta(H_0)+\left|\cov_\beta(H_\Delta,H)\right|.
\end{equation}
The final term quantifies the imbalance between the two local responses. Under exchange symmetry it vanishes, and Eq.~\eqref{Eq:simultaneous-anomalies} reduces to $-\cov_\beta(H_0,H_{\rm int}) > \var_\beta  H_0$, which remains a necessary condition for simultaneous anomalies even without the exchange symmetry. 
Eq.~\eqref{Eq:simultaneous-anomalies} also gives the necessary hierarchy $\var_\beta(H_{\rm int})>-\cov_\beta(H_0,H_{\rm int})>\var_\beta H_0+|\cov_\beta(H_\Delta,H)|$. 
Hence, at any temperature at which both subsystems lose bare energy upon heating, the interaction energy response must be sufficiently positive to preserve the non-negative response of the composite system. These relations identify the energetic mechanism behind the anomaly, but not its quantumness. 

\begin{figure*}
    \centering
    \includegraphics{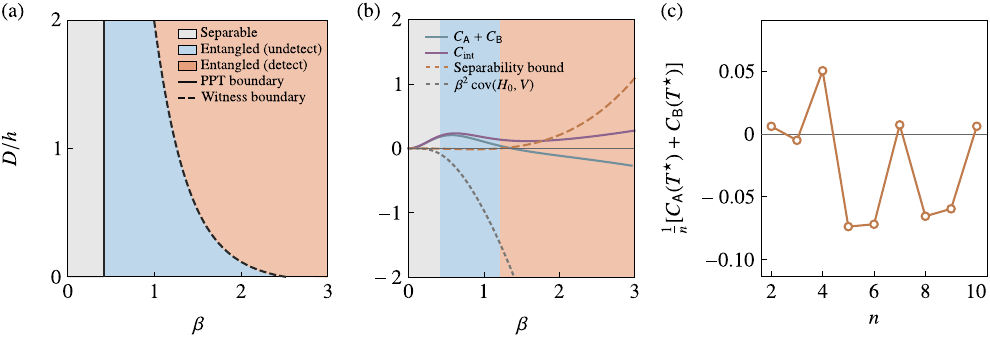}
    \caption{\emph{Thermodynamic entanglement detection in a spin chain}. (a) Detection regions for the two-spin model as functions of $\beta h$ and $D/h$, with $J/h=1$. The solid curve marks the PPT boundary~\cite{peres_96,horodecki_96}, while the dashed curve marks the violation threshold of the separability bound in Eq.~\eqref{eq:sep_bound_mt}. The gray region is separable, the blue region is NPT-entangled but not detected by the thermodynamic witness, and the orange region contains states detected by the witness. (b) heat capacity formulation for $D/h=J/h=1$. The local response $C_\A(T)+C_\B(T)$ crosses the separability bound before their sum becomes negative. The covariance contribution $\beta^2\cov_\beta(H_0,H_{\rm int})$ drives the suppression of the local response, while the positive interaction contribution $C_{\rm int}(T)$ keeps the global heat capacity non-negative. (c) Local heat capacity per spin evaluated at the witness-detection threshold $T^\star$ for the $1\mid(n-1)$ bipartition. Its sign varies with $n$, showing nonmonotonic behaviour over the system sizes considered.}
    \label{Fig:wide-plot}
\end{figure*}

\emph{\textbf{From Anomalous Local Heat Capacities to Entanglement}}. Strong energetic anti-correlations are not, by themselves, uniquely quantum. In fact, simultaneous local anomalies can occur even in commuting models whose Gibbs states are diagonal in a  product basis. Local uncertainty relations~\cite{reid_88,reid_89,hofmann,giovanetti_04,guhne_04,toth_09} are a family of separability bounds based on the fact that separable states cannot suppress the joint fluctuations of suitable sums of local observables below the limits imposed by local uncertainty relations. In particular, for a bipartite state $\rho$ on the Hilbert space $\mathcal{H}_{\A} \otimes \mathcal{H}_{\B}$, the GMVT bound~\cite{giovanetti_04} states that if $\rho \in \ms{SEP}$ i.e., $\rho = \sum_k p_k \sigma^{(\ms{A})}_k \otimes \sigma^{(\ms{B})}_k$ then $\var_\sigma(X)\var_\sigma(Y) \geq \tilde{\mathcal{O}}^2$ for $X = A_1 + B_1$ and $Y = A_2 + B_2$, where $[A_1,A_2]\neq0$, $[B_1,B_2]\neq0$ and $\tilde{\mathcal{O}} = \sum_k \tfrac{p_k}{2}(|\langle i[A_1,A_2]\rangle_{\sigma^{(\ms{A})}_k}|+|\langle i[B_1,B_2]\rangle_{\sigma^{(\ms{B})}_k}|)$. Thus, for separable states, the joint fluctuations of these global sums are bounded from below by the corresponding local uncertainty relations. The local energy $H_0=H_\ms{A}+H_\ms{B}$ is a sum of local observables whose fluctuations at equilibrium are related to the local heat capacities by $\var_\beta(H_0)=\beta^{-2}[C_\ms{A}(T)+C_\ms{B}(T)]-\cov_\beta(H_0,H_{\rm int})$. Our main result is a corresponding local uncertainty relation for local and interaction energy fluctuations, whose violation detects entanglement.
\vspace{-0cm}
\begin{res*}\label{Thm-1}
Any separable state $\rho$ of a bipartite system, with Hamiltonian $H=H_\A+H_\B + H_\text{int}$ where the interaction takes the form $H_{int}=V_\A\otimes V_\B$, obeys the uncertainty relation
\begin{align}
\var_\rho(H_0)&\var_\rho(V_\A\otimes V_\B) \geq \label{eq:sep_bound_mt} \\
    &\frac{\left(|\langle i [H_\ms{A},V_{\A}]\otimes V_{\B}\rangle_{\rho}| + |\langle V_{\A} \otimes i [H_\ms{B},V_{\B}]\rangle_{\rho}| \right)^2}{4}. \nonumber
    \end{align}
\end{res*}

\begin{sproof}
The proof begins by expanding $\var_\rho(H_0)$ and $\var_\rho(V)$ and examining which terms can be discarded for separable states. Following this, the Cauchy-Schwarz inequality is applied several times so as to arrive at a sum of covariances 
$\sum_{\ms{X} \in \{\A, \B\}}\left(\sum_k p_k\left( |\langle V_\ms{Y} \rangle_{\sigma_k}|\sqrt{|\cov_{\sigma_k}(H_\ms{X},V_\ms{X})|^2}\right)\right)^2$ which lower bounds the quantity of interest. From here, one may use the Robertson-Schr\"odinger uncertainty relation or simply $|z|^2 = \Re{z}^2 + \Im(z)^2$ for $z = \cov_{\sigma_k}(H_\ms{X},V_\ms{X})$ where the imaginary part is the commutator $-\tfrac12\langle i [H_\ms{X},V_\ms{X}]\rangle$ and the real part is the anti-commutator which we discard in the lower bound. Finally $\langle V_\ms{Y}\rangle_{\sigma_k} \langle i [H_\ms{X},V_\ms{X}]\rangle_{\sigma_k} = \langle i [H_\ms{X},V_\ms{X}] \otimes V_\ms{Y}\rangle_{\sigma_k} $ and we are able to apply $\sum p_k |\langle A_k\rangle| \geq |\sum_k p_k \langle A_k \rangle|$ by the triangle inequality to give the bound. A detailed account of the proof is provided in the SM.
\end{sproof}

Although our main result is stated for a single product interaction, the uncertainty relation depends only on the separability of $\rho$ and the product structure of a chosen observable. It therefore remains valid when $V_\A\otimes V_\B$ is one term of a more general interaction $H_{\rm int}=V_\A\otimes V_\B+W$.

We now apply our main result to a separable equilibrium state with Hamiltonian $H = H_{\A} + H_{\B} + H_\text{int}$ where the interaction $H_{\text{int}} = V_\A\otimes V_\B + W$. Expressing $\var_\beta(H_0)$ in terms of the local heat capacities then allows us to recast the uncertainty relation as
\begin{gather}
    \frac{C_\ms{A}(T) + C_\ms{B}(T)}{\beta^2} \geq \frac{\mathcal{O}_\beta^2}{4\var_\beta(V_\A\otimes V_\B)}+\cov_\beta(H_0,H_\text{int}),
\end{gather}
where $\mathcal{O}_\beta = \left(|\langle i [H_\ms{A},V_{\A}]\otimes V_{\B}\rangle_{\beta}| + |\langle V_{\A} \otimes i [H_\ms{B},V_{\B}]\rangle_{\beta}|\right)$. Consequently, the strict reverse inequality certifies that the equilibrium state is entangled. Moreover, if $\mathcal O_\beta^2/[4\var_\beta(V_\A\otimes V_\B)] +\cov_\beta(H_0,H_{\rm int})\leq0$, then any such violation implies $C_\A(T)+C_\B(T)<0$, and hence at least one local heat capacity is negative. Under the same strict violation, both local heat capacities are necessarily negative if the stronger condition 
\begin{equation}
    \frac{\mathcal{O}_\beta^2}{4\var_\beta (V_\A\otimes V_\B)} +\cov_\beta(H_0,H_{\text{int}}) \leq -|\cov_\beta(H_\Delta,H)|,
\end{equation}
holds. Under exchange symmetry, this reduces to the prior simpler condition. These relationships show that anomalous local heat capacities are neither necessary nor sufficient for the presence of entanglement, since the witness also depends on the magnitude of local and interaction energy fluctuations. On the other hand, a sufficiently negative covariance between the local and interaction energies can make a violation of separability bound imply local heat capacity anomalies. Note that the right-hand side of Eq.~\eqref{eq:sep_bound_mt} vanishes for equilibrium states of Hamiltonian with $W=\v 0$, rendering the witness trivial. Indeed, in this case $[H_\ms{X},V_{\ms X}]\otimes V_{\ms Y} = [H_\ms{X},H]$, and hence $\langle i[H_\ms{X},V_{\ms X}]\otimes V_{\ms Y} \rangle_\beta = \langle i[H_\ms{X},H]\rangle_\beta = 0$ because $[\rho_\beta,H] = 0$.

\textit{\textbf{Examples.}} These results can be illustrated by comparing two minimal models with the same local Hamiltonian $H_0=-h(\sigma_z^\A+\sigma_z^\B)$ but qualitatively different interactions. First consider the commuting Ising model $H_1=H_0-J\sigma_z^\A\sigma_z^\B$. Since all terms are diagonal in the same product basis, the Gibbs state is separable at every temperature and the witness cannot be violated. Nevertheless, anomalous local responses may still occur. For the antiferromagnetic choice $J=-2h$, both local heat capacities become negative for $\beta h>\tfrac12\ln2$. Thus, a negative local heat capacity is not by itself a signature of entanglement and can arise entirely from classical energetic correlations.

The situation changes when the interaction contains several product terms whose local factors do not commute with the corresponding local Hamiltonians. We consider the two-qubit XY model with a Dzyaloshinskii--Moriya interaction~\cite{DZYALOSHINSKY1958241,Moriya1960-xr}, $H_2=H_0-J(\sigma_x^\A\sigma_x^\B+\sigma_y^\A\sigma_y^\B)-\tfrac{D}{4}(\sigma_x^\A\sigma_y^\B-\sigma_y^\A\sigma_x^\B)$. Although the complete interaction commutes with $H_0$, its individual contributions do not. Choosing $V_\A\otimes V_\B=-(\tfrac{D}{4})\sigma_x^\A\otimes\sigma_y^\B$ and collecting the remaining terms in $W$ gives a non-trivial separability bound. For $J/h=D/h=1$, Fig.~\hyperref[Fig:wide-plot]{\ref{Fig:wide-plot}(a)} compares the witness with the exact PPT criterion~\cite{peres_96,horodecki_96}. The Gibbs state first becomes entangled at $\beta h\simeq0.43$, while the thermodynamic witness detects entanglement for $\beta h\gtrsim1.32$. As expected for a sufficient criterion, the detected region is a subset of the full entangled region. Panel~\hyperref[Fig:wide-plot]{\ref{Fig:wide-plot}(b)} shows the same transition in the heat capacity formulation. The total local response $C_\A+C_\B$ crosses the separability bound at precisely the witness threshold and changes sign only slightly later, at $\beta h\simeq1.37$. Entanglement can therefore be certified in the absence of a local heat capacity anomaly. Together with the commuting Ising example, this makes clear that anomalous local heat capacity and entanglement are distinct phenomena: the former can occur without entanglement, and the latter can be witnessed without the former. Finally, panel~\hyperref[Fig:wide-plot]{\ref{Fig:wide-plot}(c)} extends the construction to open XY--DM chains and the $1\mid(n-1)$ bipartition (see End Matter). For each displayed system size, the point shown corresponds to the first temperature at which the separability bound is violated upon cooling. The local heat capacity per spin evaluated at this threshold varies non-monotonically with $n$ and can be either positive or negative.

\emph{\textbf{Entanglement \& Positive Local Heat Capacity in Continuous Variable Systems}.} Whilst interacting quantum spins can exhibit anomalous local heat capacities, not all physical models do so. This could be a result of the correlation terms $\cov_{\beta}(H_\ms{A},H_\ms{B})$ and $\cov_{\beta}(H_\ms{A},V)$ being positive, i.e., when local energy and global energy are positively correlated, or in the anti-correlated case the negativity not being large enough to overwhelm $\var_{\beta}(H_\ms{A})$ and make $C_{\A}(T)< 0$. \textit{Are the magnitudes of these covariances and so the local heat capacity still impacted by entanglement in the absence of anomalies}? In the SM, we consider in detail the example of two coupled quantum harmonic oscillators described by $H = \frac{\omega}{2} (x^2_{\A} + x^2_{\B}) + \frac{p^2_{\A} + p^2_{\B}}{2} +\chi(x_{\A} - x_{\B})^2$, where $\omega$ characterises the local harmonic potential and $\chi$ denotes the coupling strength between the oscillators. For $H_A = \frac{1}{2}(\omega x_{\ms A}^2 + p_{\ms A}^2)$ we show that $C_{\A}(T)$ is always positive for this model by explicitly evaluating the local heat capacity using a normal mode decomposition giving

\begin{figure}[t]
    \centering
    \includegraphics{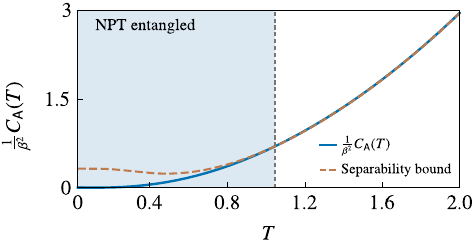}
    \caption{\emph{Local heat capacity and separability bound}. Entanglement is detected when the local heat capacity reaches the right-hand side of Eq.~\eqref{eq:boundy}. The shaded region indicates the NPT-entangled region~\cite{simon_2000,duan_2000}, given by states satisfying $\var_\beta(p_{\A} + p_{\B})\var_\beta(x_{\A} - x_{\B}) < 1$. The figure shows that violation of the bound in Eq.~\eqref{eq:boundy} detects entanglement and that entanglement constrains the local heat capacity even in the absence of anomalies. The parameter value is $\chi/\omega = 1.5$.}
    \label{fig:cv}
\end{figure}
\begin{gather}
    C_{\A}(T) = \frac{\beta^2}{16}\left(\sum_{i \in\{+,-\}}(\Omega^2_+ + \Omega^2_i)\text{csch}^2(\beta \Omega_i/2))\right),
\end{gather}
where  $\Omega_+ = \sqrt{\omega}$ and $\Omega_- = \sqrt{\omega +4\chi}$ are normal mode frequencies. Since $\beta, \Omega_\pm >0$ and $\text{csch}^2(x) > 0$ for $x >0$, it follows that $C_{\A}(T)>0$. To connect its equilibrium entanglement to the local response, we apply the PPT criterion to the covariance matrix~\cite{simon_2000,duan_2000}. For this two-mode thermal Gaussian family with $\chi\geq0$, the PPT test reduces to the product criterion~\cite{mgvt}: the state is entangled if and only if $\var_\beta(p_{\A} + p_{\B})\var_\beta(x_{\A} - x_{\B}) < 1$. Negating this condition and rearranging gives the following upper bound on the position correlation, and hence a lower bound on the effective interaction energy $-2\chi\langle x_\A x_\B\rangle_\beta$, for separable states:
\begin{gather}
\langle x_{\A}x_{\B} \rangle_\beta \leq \frac{\langle (p_{\A} + p_{\B})^2 \rangle_\beta\langle x^2_{\A} + x^2_{\B} \rangle_\beta - 1}{2\langle (p_{\A} + p_{\B})^2 \rangle_\beta} :=\mathcal{R}_\beta, \label{eq:int_energy}
\end{gather}
In the SM, we show that the local heat capacity of either quantum harmonic oscillator in a separable equilibrium state of $H$ must be lower bounded by
\begin{align}
   \hspace{-0.25cm} \frac{C_{\A}(T)}{\beta^2} \geq \var_\beta(H_\ms{A}) &+ \left(\frac{\omega(\omega + 2\chi )}{2}\right)\left(\mathcal{R}_\beta 
    -\frac{2\chi \langle x^2_{\ms A}\rangle_\beta}{\omega + 2\chi}\right)^2\nonumber\\ &+ \frac{1}{2}\langle p_\ms{A}p_{\ms{B}}\rangle_{\beta}^2 + \frac{\omega^2 \chi}{\omega + 2\chi}\langle x^2_{\ms A}\rangle^2_\beta-\frac{\chi}{4}, \label{eq:boundy}
\end{align}
for $\langle x_{\ms A} x_{\ms B} \rangle_\beta \leq \mathcal{R}_\beta \leq 2\chi\langle x_{\ms{A}}^2\rangle_\beta/(\omega+2\chi)$. This shows that, despite the absence of an anomaly, the local heat capacity is still impacted by entanglement as can be seen in Fig.~\ref{fig:cv}. 

\textbf{\textit{Heat Capacity \& Information Theoretic Quantities.}} For an interacting bipartite system with $H=H_\A+H_\B+H_{\rm{int}}$ in the Gibbs state $\rho_\beta$, we detail in the End Matter how the heat capacity can be expressed as
\begin{align}
    C_\ms{AB}(T) = C^{(0)}_{\A}(T)&+C^{(0)}_{\B}(T) -\beta^2\mathrm{\frac{d^2}{d\beta^2}}\left[D(\rho_{\A}\|\tau_{\A}) \right. \nonumber
    \\& \left. +D(\rho_{\B}\|\tau_{\B}) +\beta \langle H_\text{int}\rangle_\beta +\mathcal{I}(A:B)_\beta \right]. \label{eq:relation}
\end{align}
Here $C^{0}_{\ms X}(T) = \text{d} \text{tr}\{H_\ms{X} e^{-\beta H_\ms{X}}/Z_{\ms X}\}/\text{d}T$ is the heat capacity of the local system w.r.t a local thermal state $\tau_\ms{X}=e^{-\beta H_\ms{X}}/Z_\ms{X}$, $D(\rho\|\sigma):=\tr[\rho(\log \rho-\log\sigma)]$ is the relative entropy which is evaluated with respect to $\tau_\ms{X}$ quantifying how out-of-equilibrium or \textit{athermal}~\cite{Lostaglio_2019} the marginal is, and $\mathcal{I}(\A:\B)_\rho = S(\rho_\A)+S(\rho_\B)-S(\rho_{\A\B})$ is the mutual information. Importantly, although the relative entropies and the mutual information are non-negative, their contribution is controlled by their curvature in inverse temperature and may have either sign: contributions convex in $\beta$ suppress the response, concave ones enhance it.  The temperature dependence of athermality, the interaction energy, and correlations thus determines the extent to which the local responses are modified relative to their bare references. Since $C^{(0)}_\X(T)\geq 0$, these negative anomalies studied in this manuscript represent stark modifications. In the non-interacting limit, the global Gibbs state factorises, all terms in the bracket vanish, and additivity is recovered. Since the mutual information contains both classical and quantum correlations, this decomposition by itself does not establish a quantum origin of non-additivity. The separability criteria derived provide a more apt characterisation. 

\emph{\textbf{Discussion \& Conclusion}}
We have studied how an interacting bipartite system at equilibrium distributes its thermal response among its constituents. The local heat capacities quantify this distribution and may become negative when the bare local energies are sufficiently anti-correlated with the rest of the system. We established necessary and sufficient conditions for simultaneous anomalies and, since such anti-correlations can also occur classically, derived a separability bound on the joint fluctuations of local and interaction energies. Violating this bound certifies equilibrium entanglement using only thermal expectation values of few-body observables, without requiring Hamiltonian diagonalisation or separable-state energy bounds. For sufficiently negative covariance between local and interaction energies, the same violation also implies a negative sum of the local heat capacities. This  offers an insight into a relation between specific heat anomalies identified by Ingold et al.~\cite{ingold_09} and entanglement. Neither implies the other in general, but we identify regimes in which they necessarily coincide. See the End Matter for the explicit relationship between the local heat capacity and the specific heat capacity of~\cite{ingold_09}. The coupled-oscillator example further shows that entanglement can constrain the local thermal response even in the absence of anomalous heat capacities. Finally, the decomposition of heat capacity non-additivity attributes it to local athermality, interaction energy, and total correlations, while the quantum contribution is singled out only through separability bounds. 

Several directions merit further study. Experimentally, the local heat capacities, which are thermal susceptibilities accessible calorimetrically~\cite{PhysRevLett.133.186704, PhysRevB.99.024413} suggest the possibility of thermodynamic entanglement detection without measurements in the global energy eigenbasis. Theoretically, the freedom in splitting the interaction as $H_{\rm int}=V+W$, with the witness being trivial for $W=0$, could be optimised for a given model to obtain the tightest possible entanglement detection bound. The non-monotonic finite-size behaviour in Fig.~\hyperref[Fig:wide-plot]{\ref{Fig:wide-plot}(c)} invites a systematic study of local heat capacities across bipartitions of larger systems, in particular near criticality and phase transitions. Our work complements~\cite{Hauke2016}, where our results leave the system Hamiltonian static and detects entanglement by studying the energetic variation as temperature is changed, their work detects entanglement using the energetic response to changes in Hamiltonian parameters. \cite{deOliveiraJunior2025} can also be seen as a theoretical companion where entanglement is probed using optimal heat exchanges across quantum systems. 

\textit{Acknowledgements.} The authors thank John Goold for his hospitality, some of the work was carried out whilst the authors were visiting Trinity College Dublin. J.X. thanks Phila Rembold for introducing him to the GMVT separability bound and Ben Stratton, Mohammad Mehboudi, Mark T. Mitchison, Yuri Minoguchi, Max P. E. Lock \& especially Giuseppe Vitagliano  for useful dialogue and comments.

\textit{Tool Disclosure \& Code Availability.} The authors made use of Mathematica 14.3 and ChatGPT 5.6 Sol for symbolic analysis. The code used to generate the plots is available at~\href{https://github.com/AdeOliveiraJunior/Anomalous-local-heat capacity-and-bipartite-entanglement}{GitHub}.

\section{End Matter}

\subsection{Details of the examples}
The statement for the commuting model can be verified analytically. Setting $x=\beta h$ the partition function of $H_1$ is $Z_1 = 2(e^{\beta J} \cosh 2x + e^{-\beta J})$. For the antiferromagnetic choice $J=-2h$, the single-spin magnetisations are $\langle \sigma_z^\A\rangle_\beta = \langle \sigma_z^\B\rangle = (\cos 2x +e^{4x})^{-1}\sin 2x$. Consequently, 
\begin{equation*}
    C_A=C_B = -x^2\frac{(e^{2x}-2)(e^{2x}+1)^2} {\left[e^{4x}+\cosh(2x)\right]^2}.
\end{equation*}
Thus, $C_A=C_B<0$ when $x>\tfrac12\ln 2$.

For the XY--DM model, consider the interaction channel $O_{\A\B}=\sigma_x^\A\sigma_y^\B$. The prefactor $-\tfrac{D}{4}$ may be omitted because it cancels between the two sides of the variance inequality. Every state separable across $\A\mid\B$ satisfies
\begin{equation*}
\var_\beta(H_0)\,\var_\beta(O_{\A\B}) \geq h^2\left( \left|\langle\sigma_y^\A\sigma_y^\B\rangle_\beta\right| + \left|\langle\sigma_x^\A\sigma_x^\B\rangle_\beta\right| \right)^2.
\end{equation*}
Using $C_A+C_B=\beta^2\operatorname{cov}_\beta(H_0,H_2)$, the same condition takes the heat capacity form 
\begin{equation*}
    \frac{C_A+C_B}{\beta^2} \geq \frac{ h^2\left( \left|\langle\sigma_y^\A\sigma_y^\B\rangle_\beta\right| + \left|\langle\sigma_x^\A\sigma_x^\B\rangle_\beta\right| \right)^2 }{ \var_\beta(O_{\A\B}) } + \operatorname{cov}_\beta(H_0,H_2-H_0).
\end{equation*}
The exact two-qubit boundary shown in Fig.~\hyperref[Fig:wide-plot]{\ref{Fig:wide-plot}(a)} is obtained independently from the point at which the smallest eigenvalue of the partially transposed Gibbs state vanishes.

Finally, for the open-chain calculation, we use
\begin{equation*}
H_n = H_0 - \sum_{i=1}^{n-1} \left[ J\left( \sigma_i^x\sigma_{i+1}^x + \sigma_i^y\sigma_{i+1}^y \right) + \frac{D}{4} \left( \sigma_i^x\sigma_{i+1}^y - \sigma_i^y\sigma_{i+1}^x \right) \right].
\end{equation*}
where $H_0 = -h\sum_{i=1}^{n}\sigma_i^z$. For the bipartition $k\mid(n-k)$, the bound is applied to the channel $O_k=\sigma_k^x\sigma_{k+1}^y$ crossing the cut. Panel~\hyperref[Fig:wide-plot]{\ref{Fig:wide-plot}(c)} uses $k=1$. For each $n$, the reported temperature is the first zero of the corresponding witness function encountered upon cooling.

\subsection{Heat Capacity \& Information Theoretic Quantities}
For a fixed Hamiltonian $H_{\A\B} = H_\A+H_\B+H_{\text{int}}$, the equilibrium free energy can be decomposed into local contributions, interaction energy, and correlations:
\begin{align}
    F_\ms{AB} &= E_\ms{AB} - \frac{1}{\beta}S_\ms{AB} = \tr(\rho_\ms{AB} H_\ms{AB}) - \frac{1}{\beta}S_\ms{AB} \nonumber \\
    &=\tr[\rho_\ms{AB}(H_\ms{A}+H_{\ms{B}}+H_\text{int})] - \frac{1}{\beta}S_\ms{AB} \nonumber \\
    &= E_\ms{A} + E_{\ms{B}}+\langle H_\text{int}\rangle_{\rho_\ms{AB}}-\frac{1}{\beta}S_\ms{AB} \nonumber \\
    &= \qty(E_\ms{A} - \frac{1}{\beta}S_\ms{A}) +  \qty(E_{\ms{B}} - \frac{1}{\beta}S_{\ms{B}}) \nonumber\\
    &+ \langle H_\text{int}\rangle_{\rho_\ms{AB}}+\frac{1}{\beta}(S_\ms{A}+S_{\ms{B}}-S_\ms{AB}) \nonumber \\&=\mathcal{F}_\ms{A} + \mathcal{F}_{\ms{B}} +\langle H_\text{int}\rangle_{\rho_\ms{AB}} + \frac{1}{\beta}\mathcal{I}(A:B)_{\rho_\ms{AB}}.
\end{align}
Here, $\mathcal{F}_X := \Tr(\rho_X H_\ms{X})-\frac{1}{\beta}S(\rho_X)$ with $X\in\{\A, \B\}$. Differently from $F_\ms{AB}$, $\mathcal{F}_X$ is a nonequilibrium free energy as the marginal states might be out of equilibrium. Since $C_\ms{AB} = -\beta^2 \text{d}^2 \beta F_\ms{AB}/\text{d}\beta^2$ we can then further write
\begin{align}
    C_\ms{AB}(T) = &-\beta^2\frac{\textrm{d}^2}{\textrm{d}\beta^2}(\beta \mathcal{F}_\ms{A}) -\beta^2\frac{\textrm{d}^2}{\textrm{d}\beta^2}(\beta \mathcal{F}_{\ms{B}})\nonumber\\
    &-\beta^2\frac{\textrm{d}^2}{\textrm{d}\beta^2}(\beta H_\text{int}) - \beta^2\frac{\textrm{d}^2}{\textrm{d}\beta^2}\mathcal{I}(A:B)_\rho.
\end{align}
To separate the genuine thermal part from the nonequilibrium correction is to introduce the local Gibbs states $\tau_X \propto e^{-\beta H_\ms{X}}$ and use $\mathcal{F}_X(\rho_X) = F_X(\beta)+\beta^{-1}D(\rho_X\|\tau_X)$, so that
\begin{equation}
    -\beta^2 \frac{\textrm{d}^2}{\textrm{d}\beta^2}(\beta \mathcal{F}_X) = C^{(0)}_X(T) -\beta^2 \frac{\textrm{d}^2}{\textrm{d}\beta^2}D(\rho_X\|\tau_X), 
\end{equation}
where $C^{(0)}_X(T):= -\beta^2 \frac{\textrm{d}^2}{\textrm{d}\beta^2}(\beta F_X)$. Therefore,
\begin{align}
    C_\ms{AB}(T) = &C^{(0)}_\ms{A}(T)+C^{(0)}_{\ms{B}}(T) -\beta^2\frac{\textrm{d}^2}{\textrm{d}\beta^2}\left[D(\rho_\ms{A}\|\tau_\ms{A})\right. \nonumber\\
    &\left.  +D(\rho_{\ms{B}}\|\tau_{\ms{B}}) +\beta \langle H_\text{int}\rangle+\mathcal{I}(A:B)_\rho\right].
\end{align}
\subsection{Specific Heat Capacity using the Hamiltonian of Mean Force} In~\cite{ingold_09} specific heat capacity anomalies were investigated for one of the first times in interacting quantum systems, in particular in open quantum systems using the Hamiltonian of mean force formalism~\cite{talkner_rmp}. Here the energetics experienced by $A$ at equilibrium in a bipartite interacting system with Hamiltonian $H_\ms{AB} = H_\ms{A} + H_{\ms{B}} + V$ is described via the partition function $Z^*_\ms{A} := Z_\ms{AB}/Z_{\ms{B}}$ where $Z_{\ms{B}} = \text{tr}\{e^{-\beta H_{\ms{B}}}\}.$ The heat capacity of $A$ is found via a derivative of the free energy $C^*_\ms{A}(T) = \beta^2\dderiv{2}{}{\beta}(\ln Z^*_\ms{A}) = C_\ms{AB}(T) - C^{(0)}_{\ms{B}}(T)$ where $C^{(0)}_{\ms{B}}(T) = -\beta^2\deriv{\text{tr}\{H_{\ms{B}} e^{-\beta H_{\ms{B}}}/Z_{\ms{B}}\}}{\beta}$. This approach was used in~\cite{ingold_09} to examine specific heat anomalies in two models numerically. This quantity connects to the local heat capacity we have been considering as 
\begin{gather}
    C^{*}_\ms{A}(T) = C_\ms{A}(T) + C_\ms{B}(T) + C_\text{Int}(T) - C^{(0)}_{\ms{B}}(T),
\end{gather}
where we see that the specific heat anomaly derived from the Hamiltonian of mean force contains both local heat capacities, the interaction response and takes their difference to the bare response. From here we see that since $C(T) = C_\ms{A}(T) + C_\ms{B}(T) + C_\text{Int}(T)>0$ then if $C_\ms{A}(T) + C_\ms{B}(T)<0$ for $C^{*}_\ms{A}(T)<0$ we need 
\begin{gather}
    C^{(0)}_{\ms{B}}(T) -  (C_\ms{A}(T) + C_\ms{B}(T))>C_\text{Int}(T),
\end{gather}
since $C^{(0)}_{\ms{B}}(T) >0$ always and $C_\text{Int}(T) >0$ is positive when $C_\ms{A}(T) + C_\ms{B}(T)<0$. Therefore local heat capacity anomalies and specific heat capacities as defined in~\cite{ingold_09} are related but do not imply eachother.
\clearpage
\onecolumngrid
\section{Supplementary Material}

\subsection{Local Heat Capacity Anomaly in a Two-qubit XY chain}
Take the excitation-preserving chain
\begin{equation}
    H_{XY} = -h(\sigma^A_z+\sigma^B_z) - J(\sigma^A_x\sigma_x^B+\sigma_y^A\sigma_y^B),
\end{equation}
where $H_\ms{X}=-h\sigma_z^{\X}$ and set $J=2h$. With $x:=\beta h$ one finds the closed forms:
\begin{align}
    \operatorname{var}(H_\ms{A}) &= \frac{1}{2}h^2(2-\cosh 2x + \cosh 6x) \operatorname{sech}^2 3x ,\\ 
    \operatorname{cov}(H_\ms{A},H_{\ms{B}}+V) &= -2 h^2 (4 \cosh 2x+\cosh4x+1) \operatorname{sech}^2 3 x  \sinh ^2x, \\
\operatorname{var}(H_\ms{A}) +\operatorname{cov}(H_\ms{A},H_{\ms{B}}+V) &=  -h^2 (\cosh 4x-2 \cosh 2x) \operatorname{sech}^2 3 x \label{Eq:app-entanglement-condition}.
\end{align}
Consequently, anomalous local heat capacity appears for
\begin{equation}
    x^\star > \frac{1}{2}\text{arccosh} \qty(\frac{1+\sqrt{3}}{2}).
\end{equation}
The anomaly is a low-temperature effect present for $\beta h \gtrapprox 0.416$ and absent at high temperature $\beta h \to 0$ where fluctuations dominate and $C_\ms{A}(T) >0$.
\subsection{Entanglement Detection via a Local Uncertainty Relation of Local \& Interaction Energy}

Note that the quantity we are bounding here is present in the heat capacity squared
\begin{align}
\frac{C(T)^2}{\beta^4} &= \left(\var_\beta(H_0)+ \var_\beta(V) + 2\cov_\beta(H_0,V)\right)^2\\
&= \var_\beta(H_0)^2+ \var_\beta(V)^2 + 4\cov_\beta(H_0,V)^2 + 2\var_\beta(H_0)\var_\beta(V) + 4(\var_\beta(H_0)+ \var_\beta(V))\cov_\beta(H_0,V).
\end{align}

Inspired by the GMVT separability detection bound~\cite{giovanetti_04} we derive here a local uncertainty relation for separable states using the fluctuations of local and interaction energy expectations for bipartite interacting with $H = H_\ms{A} +H_\ms{B} +V_{\A}\otimes V_{\B}$. Assume the system to be in separable state $\rho$ admitting some decomposition $\rho = \sum_k p_k \sigma^{(\ms{A})}_k \otimes \sigma^{(\ms{B})}_k$ where $\sigma_k :=\sigma^{(\ms{A})}_k \otimes \sigma^{(\ms{B})}_k $ and consider the product of variances $\var_\rho(H_\ms{A} +H_\ms{B})\var_\rho(V_{\A} \otimes V_{\B})$. For the first term we have the simplification
\begin{align}
    \var_\rho(H_\ms{A} +H_\ms{B}) &= \var_\rho(H_\ms{A}) + \var_\rho(H_\ms{B}) +2 \cov_\rho(H_\ms{A},H_\ms{B}) \\
    &\geq \sum_k p_k\left(\var_{\sigma_k}(H_\ms{A}) + \var_{\sigma_k}(H_\ms{B})\right),
\end{align}
where we have $\cov_{\sigma_k}(H_\ms{A},H_\ms{B})=0$ for each $\sigma_k$ in the separable decomposition of $\rho$ and we have dropped the non-negative $\sum_k p_k \langle H_\ms{A} + H_\ms{B} \rangle^2_{\sigma_k} - \langle H_\ms{A} + H_\ms{B} \rangle^2_\rho = \sum_k p_k\left(\langle H_\ms{A} + H_\ms{B} \rangle_{\sigma_k} - \langle H_\ms{A} + H_\ms{B} \rangle_\rho\right)^2$ term. For the second contribution to the product we find 
\begin{align}
    \var_\rho(V_{\A} \otimes V_{\B}) &= \langle V^2_{\A} \otimes V^2_{\B} \rangle_\rho - \langle V_{\A} \otimes V_{\B} \rangle^2_\rho \\
   &= \sum_k p_k\langle V^2_{\A}\rangle_{\sigma^{(\ms{A})}_k}\langle V^2_{\B}\rangle_{\sigma^{(\ms{B})}_k} - \left( \sum_k p_k \langle V_{\A}\rangle_{\sigma^{(\ms{A})}_k}\langle V_{\B}\rangle_{\sigma^{(\ms{B})}_k}\right)^2
\intertext{now applying the flipped Cauchy-Schwarz inequality $-\left(\sum_n x^2_n\right)\left(\sum_n y^2_n\right) \leq -\left(\sum_n x_ny_n\right)^2$}
   &\geq \sum_k p_k\left(\left(\var_{\sigma^{(\ms{A})}_k}(V_{\A}) + \langle V_{\A}\rangle^2_{\sigma^{(\ms{A})}_k}\right)\left(\var_{\sigma^{(\ms{B})}_k}(V_{\B}) + \langle V_{\B}\rangle^2_{\sigma^{(\ms{B})}_k}\right) - \langle V_{\A}\rangle^2_{\sigma^{(\ms{A})}_k}\langle V_{\B}\rangle^2_{\sigma^{(\ms{B})}_k}\right)\\
   &\geq  \sum_k p_k\left(\langle V_{\B}\rangle^2_{\sigma^{(\ms{B})}_k} \var_{\sigma^{(\ms{A})}_k}(V_{\A}) + \langle V_{\A}\rangle^2_{\sigma^{(\ms{A})}_k} \var_{\sigma^{(\ms{B})}_k}(V_{\B}) \right),
\end{align}
where we dropped a positive term from the expansion for simplification. Returning to the product of interest and simplifying notation we now have 
\begin{align}
    \var_\rho(H_\ms{A} +H_\ms{B})\var_\rho(V_{\A} \otimes V_{\B}) &\geq \left(\sum_k p_k \left(\var_{\sigma_k}(H_\ms{A}) + \var_{\sigma_k}(H_\ms{B})\right)\right)\left(\sum_k p_k\left(\langle V_{\B}\rangle^2_{\sigma_k} \var_{\sigma_k}(V_{\A}) + \langle V_{\A}\rangle^2_{\sigma_k} \var_{\sigma_k}(V_{\B}) \right)\right)
\intertext{now applying the Cauchy-Schwarz inequality $\left(\sum_n x^2_n\right)\left(\sum_n y^2_n\right) \geq \left(\sum_n x_ny_n\right)^2$}
    &\geq \left(\sum_k p_k \sqrt{\left(\var_{\sigma_k}(H_\ms{A}) + \var_{\sigma_k}(H_\ms{B})\right)\left(\langle V_{\B}\rangle^2_{\sigma_k} \var_{\sigma_k}(V_{\A}) + \langle V_{\A}\rangle^2_{\sigma_k} \var_{\sigma_k}(V_{\B}) \right)}\right)^2\\
    &\geq \left(\sum_k p_k\left( |\langle V_{\B} \rangle_{\sigma_k}|\sqrt{\var_{\sigma_k}(H_\ms{A})\var_{\sigma_k}(V_{\A})}+|\langle V_{\A} \rangle_{\sigma_k}|\sqrt{\var_{\sigma_k}(H_\ms{B})\var_{\sigma_k}(V_{\B})}\right)\right)^2
\intertext{where we have applied the Cauchy - Schwarz inequality $|a||b| \geq |a\cdot b|$ to the vectors $a_k = \left(\sqrt{\var_{\sigma_k}(H_\ms{A})},\sqrt{\var_{\sigma_k}(H_\ms{B})}\right)$ and $b_k = \left(|\langle V_{\B} \rangle|_{\sigma_k}\sqrt{\var_{\sigma_k}(V_{\A})},|\langle V_{\A} \rangle|_{\sigma_k}\sqrt{\var_{\sigma_k}(V_{\B})}\right)$. Next we apply the Cauchy-Schwarz inequality to the products of variances of each term in the sum obtaining}
 &\geq \left(\sum_k p_k\left( |\langle V_{\B} \rangle_{\sigma_k}|\sqrt{|\cov_{\sigma_k}(H_\ms{A},V_{\A})|^2}+|\langle V_{\A} \rangle_{\sigma_k}|\sqrt{|\cov_{\sigma_k}(H_\ms{B},V_{\B})|^2}\right)\right)^2.
\end{align}
But now consider that 
\begin{gather}
    \frac{\cov(X,Y) - \cov(X,Y)^*}{2i} = \Im(\cov(X,Y))=\frac{\langle XY\rangle - \langle YX \rangle}{2i} = -\frac{ \langle i [X,Y]\rangle}{2}
\end{gather}
where we have used $\Im(z) = (z - z^*)/2i$ and now $|z|^2 = \Re{z}^2 + \Im(z)^2$ 
\begin{align}
    |\cov_{\sigma_k}(H_\ms{X},V_\ms{X})|^2&= \Re(\cov_{\sigma_k}(H_\ms{X},V_\ms{X}))^2 +\frac{\langle i [H_\ms{X},V_\ms{X}]\rangle_{\sigma_k}^2}{4} \geq \frac{\langle i [H_\ms{X},V_\ms{X}]\rangle_{\sigma_k}^2}{4},
\end{align}
which finally gives the separability bound
\begin{gather}\var_\rho(H_\ms{A} +H_\ms{B})\var_\rho(V_{\A} \otimes V_{\B}) \geq \left(\sum_k \frac{p_k}{2}\left( |\langle V_{\B} \rangle_{\sigma_k}\langle i [H_\ms{A},V_{\A}]\rangle_{\sigma_k}| + |\langle V_{\A} \rangle_{\sigma_k}\langle i [H_\ms{B},V_{\B}]\rangle_{\sigma_k}|\right)\right)^2 \label{eq:fin_bound}
\end{gather}
satisfied by any $\rho \in \ms{SEP}$. Absorbing the interaction expectation into the expectation of the commutator by distributivity of the tensor product i.e., $\langle V_{\B}\rangle_{\sigma_k} \langle i [H_\ms{A},V_{\A}]\rangle_{\sigma_k} = \langle i [H_\ms{A},V_{\A}] \otimes V_{\B}\rangle_{\sigma_k} $ we are able to apply $\sum p_k |\langle A_k\rangle| \geq |\sum_k p_k \langle A_k \rangle|$ by the triangle inequality we obtain the weaker bound which may be evaluated directly over the state i.e., 
\begin{gather}\var_\rho(H_\ms{A} +H_\ms{B})\var_\rho(V_{\A} \otimes V_{\B}) \geq  \frac{\left(|\langle i [H_\ms{A},V_{\A}]\otimes V_{\B}\rangle_{\rho}| + |\langle V_{\A} \otimes i [H_\ms{B},V_{\B}]\rangle_{\rho}| \right)^2}{4}, \label{eq:sep_bound}
\end{gather}
which is of particular interest for $\rho=\rho_\beta$ an equilibrium state of $H = H_\ms{A} +H_\ms{B} + V_{\A}\otimes V_{\B} + W_{\A}\otimes W_{\B}$ where $H_0 = H_\ms{A} +H_\ms{B}, X = V_{\A}\otimes V_{\B} +  W_{\A}\otimes W_{\B}$ and making use of the heat capacities defined earlier in the manuscript and $\cov_\beta(H_0,X) = \cov_\beta(X,H_0) $ we may re-express the LHS using $\var_\beta(H_0) = \beta^{-2}(C_{\A}(T)+C_{\B}(T))-\cov_\beta(H_0,X)$ we obtain
\begin{gather}
    C_{\A}(T) + C_{\B}(T) < \frac{\beta^2\left(|\langle i [H_\ms{A},V_{\A}]\otimes V_{\B}\rangle_{\beta}| + |\langle V_{\A} \otimes i [H_\ms{B},V_{\B}]\rangle_{\beta}|\right)^2}{4\var_\beta(V_{\A}\otimes V_{\B})}+\beta^2\cov_\beta(H_0,X), \label{eq:loc_heat_cap_ent_bound}
\end{gather}
which is a sufficient condition for entanglement. From this constraint we also surmise that anomalous local heat capacities can occur when $\cov_\beta(H_0,X) < 0$, so local and interaction energy expectations are anti-correlated, and is larger than the first contribution which can occur when interaction energy fluctuations are large $\var_\beta(V_{\A}\otimes V_{\B}) g\gg 1$ suppressing the first term. Note that Eq.~\eqref{eq:sep_bound} is a trivial bound for equilibrium states of Hamiltonians with a single interaction channel $H = H_0 + V$ as since $[H_\ms{A},V_{\A}]\otimes V_{\B} = [H_\ms{A},H]$ in this case we then have $\langle i[H_\ms{A},V_{\A}]\otimes V_{\B} \rangle_\beta = \langle i[H_\ms{A},H]\rangle_\beta = 0$ due to $[\rho_\beta,H] = 0$. 
\subsection{Thermodynamics \& Entanglement in two coupled Quantum Harmonic Oscillators}
\subsubsection{Introducting the Model \& Entanglement via PPT}
To better understand our results let us explore them in a well-studied, analytically solvable yet experimentally relevant example, a chain of nearest neighbour coupled quantum harmonic oscillators~\cite{eisert_audenaert_02,anders_08,AndersWinter2008}. Consider the case of two quantum harmonic oscillators described via the Hamiltonian
\begin{align}\label{eq:quad_ham}
    H_\text{QHO} &= \frac{\omega}{2} (x^2_\ms{A} + x^2_{\ms{B}}) + \frac{p^2_\ms{A} + p^2_{\ms{B}}}{2} +\chi(x_\ms{A} - x_{\ms{B}})^2 = \frac{(\omega + 2\chi)}{2}(x^2_\ms{A} + x^2_{\ms{B}}) + \frac{p^2_\ms{A} + p^2_{\ms{B}}}{2}-2\chi x_\ms{A}x_{\ms{B}} \nonumber
\end{align}
where $x_j$ and $p_j$ are the position and momentum of each mode, $\omega$ is a frequency and $\chi$ is a constant describing the coupling of the modes. Since this Hamiltonian is quadratic its thermal state is a Gaussian state and admits a complete description via the covariance matrix $\sigma_\beta$ with entries
\begin{gather}
    \sigma_{\beta_{i,j}} = \frac{1}{2}\langle R_i R_j + R_j R_i\rangle_{\beta} - \langle R_i \rangle_\beta\langle R_j \rangle_\beta
\end{gather}
and $R = (x_\ms{A}, x_{\ms{B}}, p_\ms{A}, p_{\ms{B}})$ is the vector of canonical operators. To diagonalise Eq.~\eqref{eq:quad_ham} consider the normal mode transformation 
\begin{align}
    x_\pm = \frac{x_\ms{A} \pm x_{\ms{B}}}{\sqrt{2}}\quad ,\quad   p_\pm = \frac{p_\ms{A} \pm p_{\ms{B}}}{\sqrt{2}}
\end{align}
allowing us to rewrite $H$ as 
\begin{align}
    H &= \frac{(\omega + 2\chi)}{2}\left(x^2_+ + x^2_-\right) + \frac{p^2_+ + p^2_-}{2} - \chi\left(x^2_+ - x^2_-\right)= \frac{\omega}{2}x^2_+ +\left(\frac{\omega + 4\chi}{2}\right)x^2_- + \frac{p^2_+ + p^2_-}{2}= \frac{\Omega_{\pm}}{2}\left(X^2_\pm + P^2_\pm\right) \label{eq:ham}
\end{align}
where $\Omega_+ = \sqrt{\omega}$, $\Omega_- = \sqrt{\omega + 4\chi}$ are the normal mode frequencies and we have rescaled canonical coordinates $X_\pm = x_\pm \sqrt{\Omega_\pm}$ while $P_\pm = p_\pm/\sqrt{\Omega_\pm}$. This allows us to express the thermal state of this Hamiltonian in the normal mode basis as 
\begin{gather}
    \rho_{\beta} = \frac{e^{-\beta(H_+ + H_-)}}{Z} = \frac{e^{-\beta H_+}}{Z_+} \otimes \frac{e^{-\beta H_-}}{Z_-} = \rho_{\beta_+} \otimes \rho_{\beta_-},
\end{gather}
in particular we can define annihilation and creation operators for the normal modes
\begin{align}
    a_\pm = \frac{X_\pm+iP_\pm^\dagger}{\sqrt{2}} \quad , \quad a^\dagger_\pm = \frac{X_\pm-iP_\pm^\dagger}{\sqrt{2}},
\end{align}
which using $[X,P]= i \mathbbm{1}$ satisfy $\left[a,a^\dagger\right] = \mathbbm{1}$ and so 
\begin{align}
    X_\pm = \frac{a_\pm+a_\pm^\dagger}{\sqrt{2}} \quad , \quad P^\dagger_\pm = -i\left(\frac{a_\pm-a_\pm^\dagger}{\sqrt{2}}\right)
\end{align}
which allows us to re-express the Hamiltonian Eq.~\eqref{eq:ham} as $H = \Omega_\pm\left(\frac{\mathbbm{1}}{2} + a_\pm^\dagger a_\pm\right)$. With this form in hand, we may calculate the covariance matrix $\sigma_\beta$ in the normal mode basis can easily be found. Note that due to the separability of $\rho_\beta$ in the normal mode basis we have that the covariance matrix is block diagonal in this basis $\sigma_\beta = \sigma_{\beta_+} \oplus \sigma_{\beta_-}$ where
\begin{gather}
    \sigma_{\beta_\pm} = \begin{pmatrix}
        \var_{\beta_\pm}(X_\pm) & \cov_{\beta_\pm}(X_\pm, P_\pm) \\
        \cov_{\beta_\pm}(X_\pm, P_\pm) & \var_{\beta_\pm}(P_\pm)
    \end{pmatrix},
\end{gather}
and the off-diagonal terms are equal by definition due to symmetrisation. Calculating explicit terms first note that $\langle X_\pm\rangle_{\beta_\pm} = \langle P_ \pm\rangle_{\beta_\pm} = 0$ since $\rho_{\beta_{\pm}}$ is diagonal in the basis of $H_\pm$ which contains no $a^\dagger$ or $a$ terms. Then for the off-diagonal terms the commutator $[X_\pm,P_\pm] = i\mathbbm{1}$ giving $\langle P_\pm X_\pm \rangle_{\beta_\pm} = -i +\langle X_\pm P_\pm\rangle_{\beta_\pm}$ so that 
\begin{align}
    \langle X_\pm P_\pm \rangle_{\beta_\pm} + \langle P_\pm X_\pm \rangle_{\beta_\pm} = 2\langle X_\pm P_\pm \rangle_{\beta_\pm}-i = 0
\end{align}
where we use the fact that $\langle X_\pm P_\pm \rangle_{\beta_\pm} = i/2$ for this case, giving that $\cov_{\beta_\pm}(X_\pm, P_\pm) = 0$. Next for the diagonal terms, we need only calculate $\langle X_\pm^2 \rangle_{\beta_\pm}, \langle P_\pm^2 \rangle_{\beta_\pm}$ as $\langle X_ \pm\rangle_{\beta_\pm} = \langle P_\pm\rangle_{\beta_\pm} = 0$ so consider 
\begin{align}
    \langle X_\pm^2 \rangle_{\beta_\pm} &= \left\langle \left(\frac{a_\pm +a_\pm^\dagger}{\sqrt{2}}\right)^2 \right\rangle_{\beta_\pm}  =\left\langle \frac{a_\pm^2 + {a_\pm^\dagger}^2 + 2a_\pm^\dagger a_\pm +\mathbbm{1}}{2} \right\rangle_{\beta_\pm}  = \langle n_{\pm}\rangle_{\beta_\pm} +\frac{1}{2}= \frac{\coth\left(\beta \Omega_\pm / 2\right)}{2}
\end{align}
where $n_\pm = a^\dagger a$ is the number operator in this basis and we have used the expression for the thermal occupation of a bosonic mode. Since $\langle P_\pm^2 \rangle_{\beta_\pm} =  \left\langle \left(-i\frac{a_\pm -a_\pm^\dagger}{\sqrt{2}}\right)^2 \right\rangle_{\beta_\pm}  = -\left\langle \frac{a_\pm^2 + {a_\pm^\dagger}^2  -(2a^\dagger_\pm a_\pm + \mathbbm{1})}{2} \right\rangle_{\beta_\pm} = \langle X_\pm^2 \rangle_{\beta_\pm}$ we then have that the diagonal entries are equal in each block matrix that is 
$\var_{\beta_\pm}(X_\pm) = \var_{\beta_\pm}(P_\pm) = \frac{\coth\left(\beta \Omega_\pm / 2\right)}{2}$. To investigate entanglement in this state we now rescale the coordinates and symplectically transform the covariance matrix back to the canonical basis so that carrying out a partial transpose will identify entanglement across the two modes. The covariance matrix in the $(x_\ms{A},p_\ms{A},x_{\ms{B}},p_{\ms{B}})$ basis after these transformations is then in the form
\begin{gather}
    {\sigma'}^{\text{Tr}_{\ms{B}}}_{\beta} = \begin{pmatrix}
       a_+ & 0 & a_- & 0 \\
       0 & b_+ & 0 & b_- \\
       a_- & 0 & a_+ & 0 \\
       0 & b_- & 0 & b_+
    \end{pmatrix},
\end{gather}
where 
\begin{align}
    a_\pm = \frac{1}{2}\left(\frac{\langle X_+^2 \rangle_{\beta}}{\Omega_+} \pm \frac{\langle X_-^2 \rangle_{\beta}}{\Omega_-} \right) \quad , \quad    b_\pm = \frac{\Omega_-\langle X_-^2 \rangle_{\beta} \pm \Omega_+\langle X_+^2 \rangle_{\beta}}{2}.
\end{align}
The symplectic eigenvalues of the partially transposed state can then be found using the characteristic polynomial
\begin{gather}
\text{det}\left(\lambda\mathbbm{1} -  i\mathbb{J}{\sigma'}^{\text{Tr}_{\ms{B}}}_{\beta}\right) = 0,
\end{gather}
where $\mathbb{J}$ is the symplectic form giving
\begin{align}
    \lambda_{1_\pm} = \pm \sqrt{{\frac{\Omega_-}{\Omega_+}}\langle X_+^2 \rangle_{\beta}\langle X_-^2 \rangle_{\beta}} \quad , \quad \lambda_{2_\pm} = \pm \sqrt{{\frac{\Omega_+}{\Omega_-}}\langle X_+^2 \rangle_{\beta}\langle X_-^2 \rangle_{\beta}},
\end{align}
where the smallest positive eigenvalue is $\lambda_{2_+}$ for $\Omega_+ < \Omega_-$. The PPT criterion for two mode systems states that separable Gaussians states have covariance matrix $\eta$ such that after partial transposition we still have
\begin{gather}
    \sigma'^{\text{Tr}_{\ms{B}}}_\beta + \frac{i}{2}\mathbb{J} \geq 0,
\end{gather}
implying that all symplectic eigenvalues satisfy $\nu \geq \frac{1}{2}$. With this we have that our state of interest $\rho_{\beta}$ is entangled when $\lambda_{2_+} <1/2$ that is $\langle X_+^2 \rangle_{\beta}\langle X_-^2 \rangle_{\beta} < \frac{\Omega_-}{4\Omega_+}$
\begin{align}
&\langle (X_+\sqrt{\Omega_+})^2 \rangle_{\beta}\langle (X_-/\sqrt{\Omega_-})^2 \rangle_{\beta} < \frac{1}{4},\nonumber \\
&=\langle p^2_+ \rangle_{\beta}\langle x^2_- \rangle_{\beta} < \frac{1}{4} \nonumber \\
&= \var_\beta(p_\ms{A}+p_{\ms{B}})\var_\beta(x_\ms{A} - x_{\ms{B}}) < 1 \label{eq:ent_cond}
\end{align}
for non-separable states, exposing the relationship between the PPT criterion and local sum uncertainty relationships.

Now since for any two observables $\var(A + B) = \var(\ms{A}) + \var(\ms{B})+2\cov(A,B)$ we find that the LHS of the entanglement criterion Eq.~\eqref{eq:ent_cond} can be re-expressed as 
\begin{align}
    \var_\beta(p_\ms{A} + p_{\ms{B}})\var_\beta(x_\ms{A} - x_{\ms{B}}) &= \left[\var_\beta(p_\ms{A}) + \var_\beta(p_{\ms{B}}) + 2\cov_\beta(p_\ms{A},p_{\ms{B}})\right]\left[\var_\beta(x_\ms{A}) + \var_\beta(x_{\ms{B}}) - 2\cov_\beta(x_\ms{A},x_{\ms{B}})\right]\\
    &= \left(\langle p^2_\ms{A}\rangle_\beta + \langle p^2_{\ms{B}}\rangle + 2\langle p_\ms{A} p_{\ms{B}}\rangle_\beta\right)\left(\langle x^2_\ms{A}\rangle_\beta + \langle x^2_{\ms{B}}\rangle_\beta - 2\langle x_\ms{A} x_{\ms{B}}\rangle_\beta\right)
\end{align}
where the 1st moments disappear since these are expectation values w.r.t a thermal state whose Hamiltonian features no first order terms. In other words, if entanglement is present we have that 
\begin{align}
  &\left(\langle (p_\ms{A} + p_{\ms{B}})^2\rangle\right)\left(\langle x^2_\ms{A}\rangle_\beta + \langle x^2_{\ms{B}}\rangle_\beta - 2\langle x_\ms{A} x_{\ms{B}}\rangle_\beta\right) < 1 \implies \langle x_\ms{A}x_{\ms{B}} \rangle_\beta > \frac{\langle (p_\ms{A} + p_{\ms{B}})^2 \rangle_\beta\langle x^2_\ms{A} + x^2_{\ms{B}} \rangle_\beta - 1}{2\langle (p_\ms{A} + p_{\ms{B}})^2 \rangle_\beta}:=\mathcal{R}_\beta \label{eq:int_bound}
\end{align}
by manipulating the PPT criterion inequality.

We now turn our attention to the local heat capacities of the two quantum harmonic oscillators and investigate their behaviour in the presence of entanglement. Consider first for $H_\ms{A} = \frac{1}{2}\left(\omega x^2_\ms{A} +p^2_\ms{A}\right)$
\begin{align}
   \deriv{\langle H_\ms{A} \rangle_\beta}{\beta} &= \deriv{\langle \frac{1}{2}\left(\omega x^2_\ms{A} +p^2_\ms{A}\right) \rangle_\beta}{\beta} = \deriv{\langle \frac{1}{4}\left(\omega(x_+ + x_-)^2 +(p_+ + p_-)^2 \right) \rangle_\beta}{\beta}\nonumber\\
    &= \deriv{\langle \frac{1}{4}\left(\omega(x^2_+ + x^2_- + 2x_+x_-) +(p^2_+ + p^2_- + 2p_+p_-)\right) \rangle_\beta}{\beta}\nonumber\\
    &= \deriv{\langle \frac{1}{4}\left(\omega(x^2_+ + x^2_-) +(p^2_+ + p^2_-)\right) \rangle_\beta}{\beta} = \deriv{\langle \frac{\Omega_+ }{4\Omega_-}(\Omega_-X^2_+ + \Omega_+X_-^2) + \frac{1}{4}(\Omega_+P_+^2 + \Omega_-P^2_-)\rangle_\beta}{\beta}\nonumber\\
    &= \deriv{}{\beta}\left( \sum_{i \in \{+,-\}}\frac{\Omega_+}{8\Omega_-}\Omega_{\bar{i}}\coth(\beta \Omega_i/2) + \frac{\Omega_i}{8}\coth(\beta \Omega_i/2)\right)
\end{align}
and $\deriv{}{x}(\coth(x)) = -\text{csch}^2(x)$ giving 
\begin{align}
    C_\ms{A}(T) &= -\beta^2   \deriv{\langle H_\ms{A} \rangle_\beta}{\beta} = \frac{\beta^2}{16}\left(\sum_{i \in\{+,-\}}\frac{\Omega_+ }{\Omega_-}\Omega_{\bar{i}}\Omega_i\text{csch}^2(\beta \Omega_i/2) + \Omega^2_i\text{csch}^2(\beta \Omega_i/2)\right) \\
    &=  \frac{\beta^2}{16}\left(\sum_{i \in\{+,-\}}(\Omega^2_+ + \Omega^2_i)\text{csch}^2(\beta \Omega_i/2))\right)
\end{align}
showing that since $\beta, \Omega_\pm >0$ and $\text{csch}^2(x) > 0$ for $x >0$ then $C_\ms{A}(T)>0$ for this model and it has no anomalous local heat capacity. With that said, as we have shown in the main text we can decompose the local heat capacity as 
\begin{gather}
    C_\ms{A}(T) = \beta^2\cov_{\beta}(H_\ms{A},H) = \beta^2\left(\var_{\beta}(H_\ms{A}) + \cov_{\beta}(H_\ms{A},H_{\ms{B}})+\cov_{\beta}(H_\ms{A},V)\right)  
\end{gather}
where we wish to see a connection with entanglement via $\cov_{\beta}(H_\ms{A},H_{\ms{B}})$ and $J\cov_{\beta}(H_\ms{A},V)$ whose magnitude may be influenced by the presence of entanglement. To examine this, let us calculate each of the three contributions to $C_\ms{A}(T)$ for which we will need the identities 
\begin{align}
    \langle R_1 R_2 R_3 \rangle &= \langle R_1 \rangle\langle R_2 R_3 \rangle + \langle R_2 \rangle\langle R_1 R_3 \rangle+ \langle R_3\rangle\langle R_1 R_2 \rangle - 2\langle  R_1\rangle\langle R_2\rangle\langle R_3 \rangle, \\
    \langle R_1 R_2 R_3 R_4 \rangle &= \langle R_1 R_2 \rangle\langle R_3 R_4 \rangle + \langle R_1 R_3\rangle \langle R_2 R_4 \rangle + \langle R_1 R_4\rangle\langle R_2 R_3\rangle ,
\end{align}
which follow from the use of normal ordering and Wick's theorem. We then have 
\begin{align}
    \var_{\beta}(H_\ms{A}) &= \left\langle \left(\frac{1}{2}\left(\omega x^2_\ms{A} +p^2_\ms{A}\right)\right)^2 \right\rangle_{\beta} - \left\langle \frac{1}{2}\left(\omega x^2_\ms{A} +p^2_\ms{A}\right)\right\rangle^2_{\beta} \nonumber\\
    &= \frac{\omega^2}{4}\left\langle x^4_\ms{A}\right\rangle_{\beta} + \frac{1}{4}\left\langle p^4_\ms{A}\right\rangle_{\beta}+ \frac{\omega}{2}\left\langle x^2_\ms{A}p^2_\ms{A}\right\rangle_{\beta} - \frac{\omega^2}{4}\langle x^2_\ms{A}\rangle^2_{\beta} -  \frac{1}{4}\langle p^2_\ms{A} \rangle^2_{\beta} - \frac{\omega}{2}\langle x^2_\ms{A}\rangle_{\beta}\langle p^2_\ms{A}\rangle_{\beta}\nonumber\\  
    &= \frac{\omega^2}{2}\left\langle x^2_\ms{A}\right\rangle_{\beta}^2 + \frac{1}{2}\left\langle p^2_\ms{A}\right\rangle_{\beta}^2 + \omega\left\langle x_\ms{A}p_\ms{A}\right\rangle_{\beta}^2 + \frac{\omega}{2}\langle x^2_\ms{A}\rangle_{\beta}\langle p^2_\ms{A}\rangle_{\beta}  - \frac{\omega}{2}\langle x^2_\ms{A}\rangle_{\beta}\langle p^2_\ms{A}\rangle_{\beta}\nonumber\\
    &= \frac{\omega^2}{2}\left\langle x^2_\ms{A}\right\rangle_{\beta}^2 + \frac{1}{2}\left\langle p^2_\ms{A}\right\rangle_{\beta}^2  - \frac{\omega}{4}  \\
    \nonumber\\
    \cov_{\beta}(H_\ms{A},H_{\ms{B}}) &=  \left\langle \left(\frac{1}{2}\left(\omega x^2_\ms{A} +p^2_\ms{A}\right)\right)\left(\frac{1}{2}\left(\omega x^2_\ms{B} +p^2_\ms{B}\right)\right)\right\rangle_{\beta} - \left\langle \frac{1}{2}\left(\omega x^2_\ms{A} +p^2_\ms{A}\right)\right\rangle_{\beta}\left\langle\frac{1}{2}\left(\omega x^2_\ms{B} +p^2_\ms{B}\right)\right\rangle_{\beta}\nonumber\\
    &=\frac{\omega^2}{4}\left(\langle x^2_\ms{A}\rangle_{\beta}\langle x^2_{\ms{B}}\rangle_{\beta} + 2\langle x_\ms{A}x_{\ms{B}}\rangle_{\beta}^2\right) + \frac{\omega}{4}\left(\langle x^2_\ms{A}\rangle_{\beta}\langle p^2_{\ms{B}}\rangle_{\beta}+\langle p^2_\ms{A}\rangle_{\beta}\langle x^2_{\ms{B}}\rangle_{\beta}\right)\nonumber\\ &+ \frac{1}{4}\left(\langle p^2_\ms{A}\rangle_{\beta}\langle p^2_{\ms{B}}\rangle_{\beta} + 2\langle p_\ms{A}p_{\ms{B}}\rangle_{\beta}^2\right) - \left\langle \frac{1}{2}\left(\omega x^2_\ms{A} +p^2_\ms{A}\right)\right\rangle_{\beta}\left\langle\frac{1}{2}\left(\omega x^2_\ms{B} +p^2_\ms{B}\right)\right\rangle_{\beta}\nonumber\\
    &=\frac{\omega^2}{2}\langle x_\ms{A}x_{\ms{B}}\rangle_{\beta}^2 + \frac{1}{2}\langle p_\ms{A}p_{\ms{B}}\rangle_{\beta}^2
    \end{align}
    \begin{align}
    \cov_{\beta}(H_\ms{A},V) &= \cov_{\beta}(H_\ms{A},\chi(x_\ms{A} - x_{\ms B})^2) = \chi\left( \cov_{\beta}(H_\ms{A},x^2_\ms{A}) + \cov_{\beta}(H_\ms{A},x^2_{\ms B})) - 2 \cov_{\beta}(H_\ms{A}, x_\ms{A}x_{\ms B})\right) 
\end{align}
where we have 
\begin{align}
     \cov_{\beta}(H_\ms{A},x^2_\ms{A}) &=\frac{1}{2}\left\langle \omega x^4_{\ms A} + p^2_{\ms A} x^2_{\ms A}\right\rangle_\beta  - \left\langle\frac{1}{2}\left(\omega x^2_\ms{A} +p^2_\ms{A}\right) \right\rangle_\beta\left\langle x^2_{\ms A}\right\rangle_\beta \nonumber \\
     &= \frac{3\omega}{2} \langle x^2_{\ms A}\rangle^2_\beta + \frac{1}{2}\langle p^2_{\ms A}\rangle_\beta\langle x^2_{\ms A}\rangle_\beta + \langle p_{\ms A}x_{\ms A}\rangle_\beta^2 - \frac{\omega}{2} \langle x^2_{\ms A}\rangle^2_\beta - \frac{1}{2}\langle p^2_{\ms A}\rangle_\beta\langle x^2_{\ms A}\rangle_\beta\nonumber\\
     &= \omega \langle x^2_{\ms A}\rangle^2_\beta - \frac{1}{4}
\end{align}
similarly 
\begin{align}
    \cov_{\beta}(H_\ms{A},x^2_\ms{B}) &=\frac{1}{2}\left\langle \omega x^2_{\ms A}x^2_{\ms B} + p^2_{\ms A} x^2_{\ms B}\right\rangle_\beta  - \left\langle\frac{1}{2}\left(\omega x^2_\ms{A} +p^2_\ms{A}\right) \right\rangle_\beta\left\langle x^2_{\ms B}\right\rangle_\beta \nonumber\\
    &= \frac{\omega}{2}\left(\langle x^2_{\ms A}\rangle_\beta\langle x^2_{\ms B}\rangle_\beta + 2 \langle x_{\ms A}x_{\ms B}\rangle_\beta^2\right) + \frac{1}{2}\left(\langle p^2_{\ms A}\rangle_\beta\langle x^2_{\ms B}\rangle_\beta + 2 \langle p_{\ms A}x_{\ms B}\rangle_\beta^2\right)  - \frac{\omega}{2} \langle x^2_{\ms A}\rangle_\beta\langle x^2_{\ms B}\rangle_\beta - \frac{1}{2}\langle p^2_{\ms A}\rangle_\beta\langle x^2_{\ms B}\rangle_\beta \nonumber\\
    &= \omega\langle x_{\ms A} x_{\ms B}\rangle^2_\beta
\end{align}
and lastly 
\begin{align}
    \cov_{\beta}(H_\ms{A},x_\ms{A}x_{\ms B}) &=\frac{1}{2}\left\langle \omega x^2_{\ms A}x_{\ms A}x_{\ms B} + p^2_{\ms A} x_{\ms A}x_{\ms B}\right\rangle_\beta  - \left\langle\frac{1}{2}\left(\omega x^2_\ms{A} +p^2_\ms{A}\right) \right\rangle_\beta\left\langle x_{\ms A}x_{\ms B}\right\rangle_\beta \nonumber\\
    &=\frac{3\omega}{2} \langle x^2_{\ms A}\rangle_\beta\langle x_{\ms A} x_{\ms B}\rangle_\beta + \frac{1}{2} \langle p^2_{\ms A}\rangle_\beta\langle x_{\ms A} x_{\ms B}\rangle^2_\beta + \cancel{2\langle p_{\ms A} x_{\ms A}\rangle_\beta \langle p_{\ms A} x_{\ms B}\rangle_\beta} - \frac{\omega}{2} \langle x^2_{\ms A}\rangle_\beta\langle x_{\ms A} x_{\ms B}\rangle_\beta - \frac{1}{2}\langle p^2_{\ms A}\rangle_\beta\langle x_{\ms A} x_{\ms B}\rangle_\beta\nonumber \\
    &= \omega\langle x^2_{\ms A}\rangle_\beta\langle x_{\ms A} x_{\ms B}\rangle_\beta
\end{align}
giving
\begin{gather}
\cov_{\beta}(H_\ms{A},V) = \omega \chi \langle x^2_{\ms A}\rangle^2_\beta - \frac{\chi}{4} + \omega\chi\langle x_{\ms A} x_{\ms B}\rangle^2_\beta -2\omega\chi\langle x^2_{\ms A}\rangle_\beta\langle x_{\ms A} x_{\ms B}\rangle_\beta.
\end{gather}
We can then express the local heat capacity as 
\begin{align}
    \frac{C_\ms{A}(T)}{\beta^2} &= \var_\beta(H_\ms{A}) + \frac{1}{2}\langle p_\ms{A}p_{\ms{B}}\rangle_{\beta}^2 + \left(\frac{\omega^2}{2}  + \omega\chi\right)\langle x_\ms{A}x_\ms{B}\rangle_\beta^2 
    -2 \omega\chi\langle x^2_\ms{A}\rangle_\beta\langle x_\ms{A}x_\ms{B}\rangle_\beta+\omega\chi\langle x^2_\ms{A}\rangle_\beta^2 - \frac{\chi}{4} \nonumber\\
    &= \var_\beta(H_\ms{A}) + \frac{1}{2}\langle p_\ms{A}p_{\ms{B}}\rangle_{\beta}^2 + \left(\frac{\omega(\omega + 2\chi )}{2}\right)\langle x_\ms{A}x_\ms{B}\rangle_\beta^2 
    -2 \omega\chi\langle x^2_{\ms A}\rangle_\beta\langle x_\ms{A}x_\ms{B}\rangle_\beta+\omega\chi\langle x^2_\ms{A}\rangle_\beta^2 - \frac{\chi}{4} \nonumber\\
    &= \var_\beta(H_\ms{A}) + \frac{1}{2}\langle p_\ms{A}p_{\ms{B}}\rangle_{\beta}^2 + \left(\frac{\omega(\omega + 2\chi )}{2}\right)\left(\langle x_\ms{A}x_\ms{B}\rangle_\beta^2 
    -\frac{4\chi \langle x^2_{\ms A}\rangle_\beta}{\omega + 2\chi}\langle x_\ms{A}x_\ms{B}\rangle_\beta\right)+\omega\chi\langle x^2_\ms{A}\rangle_\beta^2 - \frac{\chi}{4}\nonumber\\
    &= \var_\beta(H_\ms{A}) + \frac{1}{2}\langle p_\ms{A}p_{\ms{B}}\rangle_{\beta}^2 + \left(\frac{\omega(\omega + 2\chi )}{2}\right)\left(\left(\langle x_\ms{A}x_\ms{B}\rangle_\beta 
    -\frac{2\chi \langle x^2_{\ms A}\rangle_\beta}{\omega + 2\chi}\right)^2 - \frac{4\chi^2 \langle x^2_{\ms A}\rangle^2_\beta}{(\omega + 2\chi)^2}\right)+\omega\chi\langle x^2_\ms{A}\rangle_\beta^2 - \frac{\chi}{4} \nonumber\\
    &= \var_\beta(H_\ms{A}) + \frac{1}{2}\langle p_\ms{A}p_{\ms{B}}\rangle_{\beta}^2 + \left(\frac{\omega(\omega + 2\chi )}{2}\right)\left(\langle x_\ms{A}x_\ms{B}\rangle_\beta 
    -\frac{2\chi \langle x^2_{\ms A}\rangle_\beta}{\omega + 2\chi}\right)^2 + \frac{\omega \chi(\omega+2\chi) \langle x^2_{\ms A}\rangle^2_\beta - 2\omega\chi^2\langle x^2_{\ms A}\rangle^2_\beta}{\omega + 2\chi} - \frac{\chi}{4}\nonumber \\
    &= \var_\beta(H_\ms{A}) + \frac{1}{2}\langle p_\ms{A}p_{\ms{B}}\rangle_{\beta}^2 + \left(\frac{\omega(\omega + 2\chi )}{2}\right)\left(\langle x_\ms{A}x_\ms{B}\rangle_\beta 
    -\frac{2\chi \langle x^2_{\ms A}\rangle_\beta}{\omega + 2\chi}\right)^2 + \omega^2 \chi\langle x^2_{\ms A}\rangle^2_\beta - \frac{\chi}{4}
\end{align}

Making use of the bound on the interaction energy for non-separable states which we attained earlier from the PPT criterion Eq.~\eqref{eq:int_bound} we can bound the interaction contributions. In particular, if $\rho_{\beta}$ is a separable equilibrium state of $H$ then we have 

\begin{align}
    \frac{C_\ms{A}(T)}{\beta^2} \geq \var_\beta(H_\ms{A}) + \frac{1}{2}\langle p_\ms{A}p_{\ms{B}}\rangle_{\beta}^2 + \left(\frac{\omega(\omega + 2\chi )}{2}\right)\left(\mathcal{R}_\beta 
    -\frac{2\chi \langle x^2_{\ms A}\rangle_\beta}{\omega + 2\chi}\right)^2 + \frac{\omega^2 \chi}{\omega + 2\chi}\langle x^2_{\ms A}\rangle^2_\beta - \frac{\chi}{4}
\label{eq:ent_bound}
\end{align}
for separable equilibrium states of two coupled quantum harmonic oscillators when $\langle x_{\ms A} x_{\ms B} \rangle_\beta \leq \mathcal{R}_\beta \leq 2\chi\langle x_{\ms{A}}^2\rangle_\beta/(\omega+2\chi)$.

When using Eq.~\eqref{eq:ent_bound} to construct the bound plotted over the full temperature range in Fig.\ref{fig:cv}, one must account for the fact that the completed square is not globally monotonic in $\langle x_{\ms{A}}x_{\ms{B}}\rangle_\beta$. 
In particular, defining $f_\beta(y):=\left[y-\frac{2\chi \langle x^2_{\ms A}\rangle_\beta}{\omega + 2\chi}\right]^2$, its minimum occurs at $y_*:=\frac{2\chi \langle x^2_{\ms A}\rangle_\beta}{\omega + 2\chi}$. The function $f_\beta(y)$ decreases for $y<y_*$ and increases for $y>y_*$. Consequently, the direct substitution of the PPT-derived bound $\langle x_{\ms{A}}x_{\ms{B}}\rangle_\beta\leq\mathcal{R}_\beta$ for separable states is valid only while $\mathcal{R}_\beta\leq y_*$. If $\mathcal{R}_\beta>y_*$, the allowed interval already contains the minimum of the completed square, and substituting $\mathcal{R}_\beta$ would incorrectly continue the bound beyond the decreasing branch of $f_\beta(\cdot)$. We therefore introduce $\widetilde{\mathcal{R}}_\beta:=\min\{\mathcal{R}_\beta,y_*\}$ and use $\widetilde{\mathcal{R}}_\beta$ in place of $\mathcal{R}_\beta$ in the plotted bound. For $\mathcal{R}_\beta\leq y_*$ this reproduces Eq.~\eqref{eq:ent_bound}, whereas for $\mathcal{R}_\beta>y_*$ it evaluates the completed square at its global minimum, ensuring that the resulting separability bound remains valid at all temperatures and does not produce spurious violations.

\end{document}